\documentclass[amsmath,twocolumn]{revtex4}
\usepackage{graphicx}
\newcommand{\vc}{\vspace*{\fill}}
\newcommand{\np}{\newpage\vc}

\begin{document}
\bibliographystyle{prsty}

\title{Ab Initio Molecular Dynamics Study of Aqueous Solvation of Ethanol 
and Ethylene}
\author{Titus S. van Erp and Evert Jan Meijer}
\affiliation{Department of Chemical Engineering,
Universiteit van Amsterdam,
Nieuwe Achtergracht 166, NL-1018 WV AMSTERDAM, The Netherlands}

\date{November 28, 2002}

\begin{abstract}
  The structure and dynamics of aqueous solvation of ethanol and
  ethylene are studied by DFT-based Car-Parrinello molecular dynamics.
  We did not find an enhancement of the structure of the hydrogen
  bonded network of hydrating water molecules. Both ethanol and
  ethylene can easily be accommodated in the hydrogen-bonded network
  of water molecules without altering its structure. This is supports
  the conclusion from recent neutron diffraction experiments that
  there is no hydrophobic hydration around small hydrophobic groups.
  Analysis of the electronic charge distribution using Wannier
  functions shows that the dipole moment of ethanol increases from 1.8
  D to 3.1 D upon solvation, while the apolar ethylene molecule
  attains an average dipole moment of 0.5~D.  For ethylene, we
  identified configurations with $\pi$-H bonded water molecules, that
  have rare four-fold hydrogen-bonded water coordination, yielding
  instantaneous dipole moments of ethylene of up to 1~D. The results
  provide valuable information for the improvement of empirical force
  fields, and point out that for an accurate description of the
  aqueous solvation of ethanol, and even of the apolar ethylene,
  polarizable force fields are required.
\end{abstract}
\maketitle

\clearpage
\section{Introduction}
\label{secInt}

The study of the solvation of alcohols in aqueous solution is of
fundamental interest in physics, biology and chemistry, but also of
importance in technical applications
\cite{FranksIves,Eisenberg,Franks,FranksReid,Blandamer,Luck,Franksv4,Alfsen,
Franks2,FranksDes,Mohr}.  
Among the alcohols, ethanol is one the most well studied compounds.
Aqueous ethanol solutions are a common in chemical research and
industry applications.  Ethanol can be produced by aqueous hydration of
ethylene, that is readily available from natural sources.  This process
that can be accelerated by acid catalysis.
The reverse route of decomposing of ethanol into water and ethylene is
also of importance.  For example, for developing countries who do not
have a large supply of fossil fuels, dehydration of ethanol obtained
from biomass is often the most economical way to produce ethylene.  As
is well known, solvation structures play a crucial role in aqueous
solution chemistry, where reactive events often require a significant
reordering of the water molecules in the solvation shell.  The
solvation of ethanol and ethylene are therefore crucial in the course
of their (acid-catalyzed) interconversion. The aqueous solvation of
these molecules will be addressed in the present paper.  Elsewhere we
report on an {\em ab initio} molecular dynamics study of
the acid-catalyzed conversion \cite{ownfut}.

Ethanol and ethylene have distinct solvation properties in aqueous
solution.  Ethanol is easily soluble as its polar hydroxyl group can
participate in the hydrogen bonded network and the hydrophobic ethyl
group is relatively small. In contrast, the apolar
ethylene molecule has a much weaker interaction with water and is
generally considered to be hydrophobic.
Mixtures of water and ethanol have been studied extensively, both
experimentally as by molecular simulation.  Experimental studies
employing NMR \cite{Coccia,Harris,Mizuno,Ludwig,Lamanna}, ultrasonic
absorption \cite{Brai}, infrared absorption spectroscopy
\cite{Harris,Mizuno,Nishi,Onori}, mass spectroscopy
\cite{Nishi,Nishi0}, X-ray diffraction measurements
\cite{Nishi,Petrillo}, neutron diffraction
\cite{Petrillo,Turner,Sidhu}, and dielectric relaxation measurements
\cite{Mashimo,Bao,Sato,Petong} have been performed to unravel the
solvation properties of ethanol.  Molecular simulation studies using
empirical force fields have addressed the equation of state,
thermodynamics, and structure and dynamics of
solvation \cite{Mueller-Plathe,Levchuk,Fidler,Tarek,Tarek2} of aqueous
ethanol solutions.  The following general picture of the aqueous
solvation in dilute solutions has emerged: the hydroxyl group
participates in the hydrogen bonded network, while the hydrophobic
alkyl group is accommodated in the hydrogen bonded network of
water molecules.  The nature of hydration structure around the
hydrophobic part of alcohols is still a controversial subject. It has
been suggested 
\cite{FranksIves,Ludwig,Lamanna,Franks,FranksReid,Franksv4,Blandamer,FranksDes}
that hydrophobic solutes enforces the network of
hydrogen bonded water molecules around it and decreases their
mobility, a notion referred to as hydrophobic 
hydration \cite{Frank,Ben-Naim}. However,
recent experimental and computational studies have shown that for
small alcohols the structure and dynamics of the water molecules in
the solvation shell is almost identical to that of bulk water
\cite{Fidler,Turner,Sidhu}.

Much less is known for the solvation of ethylene in water under
ambient conditions.  Experimental \cite{Sugahara} and
theoretical \cite{Tanaka,Kvamme} work addressed the clathrate hydrates
of ethylene in water.
The isolated ethylene-water complex has been a
subject of various studies.  In the lowest energy configuration,
the ethylene molecule forms a weak
bond with water.  An early \emph{ab initio} study of Del Bene in 1974
\cite{DelBene} has characterized this interaction as a $\pi$-H
hydrogen bond of a water proton with the $\pi$ electrons of the
C=C bond.  Several experimental techniques have been applied to
measure the strength of this interaction, such as the matrix isolation
study of Engdahl and Nelander \cite{Engdahl,Engdahl86}, the microwave spectra
study of Andrews and Kuczkowski \cite{Andrews} and the molecular-beam
measurements of Peterson and Klemperer \cite{Peterson}.  Recently
Tarakeshwar et al.  \cite{Tarakeshwar,Tarakeshwar2,Tarakeshwar3} and
Dupr\'e and Yappert \cite{Dupre}
have performed calculations on the ethylene-water
complexes with high level \emph{ab initio} methods.
The interaction is 
weak compared to a hydrogen bonds such as in the water dimer.  The
role of the $\pi$-H bond in aqueous solvation under ambient conditions
is still an open question.

Molecular simulation provides an approach to study the microscopic
behavior of liquids complementary to experimental studies. All
molecular simulations studies of aqueous ethanol solutions up to now
are based on empirical force fields that are designed to reproduce a
selection of experimental data.  Obviously, molecular simulations
based on these potentials do not provide a picture completely
independent from experiment.  Moreover, the reliability of the results
at conditions that are significantly different from those where the
potential was designed for, may be questionable.  Density functional
theory (DFT) based molecular dynamics (MD) simulation, such as the
Car-Parrinello molecular dynamics method \cite{CaPa85}, where the
interactions are calculated by accurate electronic structure
calculations, provides a route to overcome these limitations. This has
been demonstrated in studies of liquid water
\cite{Laasonen,Sprik,SiBe97} and aqueous solvation
\cite{Marx,ErpMe01_erratum,Raugei}.  Important advantages of DFT-MD
over force-field MD are that it intrinsically incorporates
polarization, that it accounts for the intra-molecular motion and
therefore allows for a direct comparison with spectroscopy of
intra-molecular vibrations, and that it yields detailed information on
the electronic properties, such as the energy levels of electronic
states and the charge distribution. In a broader chemical perspective
it is important to note that DFT-MD is capable to study 
chemical reactions in solution, where force-field MD would fail
completely as it cannot account for the change in chemical bonding.

Here, we report on a DFT-based MD simulation of the solvation of
ethanol and ethylene.  First we describe the simulation methods. Then
we show results of geometries and energetics of relevant gas-phase
complexes, that will serve as a validation of the numerical methods
employed.  Subsequently, results of structure, dynamics, and
polarization of the solvated species will be shown. We conclude with a
discussion.

\section{Computational Methods}
\label{secmethods}
Electronic structure calculations are performed using the Kohn-Sham
\cite{KoSh65} formulation of DFT \cite{HoKo64}.  We employed the
gradient-corrected BLYP functional \cite{Beck88_2,LeYa88}, that has
proven to give a good description of the structure and dynamics of
liquid water \cite{Sprik}.  The DFT-based MD simulations of aqueous
ethanol and ethylene are performed using the Car-Parrinello method as
implemented in the CPMD package \cite{CPMD30f}. Norm-conserving
Martins-Trouillier pseudopotentials \cite{TrMa91} are used to restrict
the number of electronic states to those of the valence electrons.
Cut-off radii for H, O and C atoms were chosen to be 0.50, 1.11 and
1.23 a.u., respectively both for the l=s and l=p terms.  The Kohn-Sham
orbitals are expanded in a plane wave basis, matching the periodicity
of the periodic box with waves up to a kinetic energy of 70~Ry.  With
this basis set energies and geometries are converged within 0.25~kcal/mol
and 0.01 \AA~, respectively. Vibrational frequencies are converged
within 1 \%, except for C-O and O-H stretch modes that are
underestimated by 3 \% and 5 \% compared to the basis set limit values
\cite{ErpMe01_erratum}.  In the molecular dynamics calculations, the fictitious
mass associated with the plane-wave coefficients is chosen at 900
a.u., which allowed for a time step in the numerical integration of
the equations-of-motion of 0.145 fs.
 
To validate the computational approach we compared CPMD results for
energies and structures of relevant gas-phase molecules and complexes
with state-of-the-art atomic-orbital DFT-BLYP calculations performed
with ADF \cite{ADF2002,Velde,Fonseca,ADF_basisVII}, and other
high-level quantum chemical results taken from literature.  The
gas-phase calculations with CPMD were performed in a large periodic
box of 10 \AA~ using the screening technique of \cite{BaLa93} to
eliminate the interactions among periodic images. We have not included
zero-point energies in the energies of the gas-phase compounds. This
also holds for computed energies taken from literature and referred to in the
present paper. This, to ensure a proper comparison between our results
and those from literature.

The MD simulations of the solutions were performed for a 'small' and a
'large' system to asses the finite size effects.  For the small system
a simulation of 10 ps was performed in a cubic periodic box of length
10.07 \AA, both for an ethanol solution of 31 waters and 1 ethanol as
for an ethylene solution of 32 water and 1 ethylene.  For the large
system 5 ps simulations were performed using a periodic box with bcc
symmetry and a volume of 1977.6 \AA$^3$.  This periodic cell, a
truncated octahedron, is in shape closer to a sphere than a simple
cube, and therefore better suited for liquid simulation.  The large
ethanol solution consisted of 63 waters and 1 ethanol, while the large
ethylene solution consisted of 64 waters and 1 ethylene.  The box
sizes for both the ethanol and ethylene solutions were set to match
the experimental densities of the ethanol solutions under ambient
conditions.  For the ethylene system this will be slightly larger than
the experimental density, as the effective volume of ethanol is a bit
smaller than the combined volume of ethylene + water. However, we do
not expect this to give rise to observable changes in the calculated
properties.  For reference we performed 10~ps MD simulations of a
single ethanol and ethylene in a periodic cubic box of 10 \AA~ and a
pure water systems of 32 water molecules in a cubic box of 9.85 \AA~
for simulation times of 10 ps.  For all simulations there was an
initial equilibration trajectory of 1~ps.  Temperature was controlled
by a Nos\'e-Hoover thermostat \cite{Nose84_1,Nose84_2,Hoov85} and
fixed at 300~K.
 
\section{\bf Gas-phase complexes}
\label{secgas}
The ethanol monomer has two stable conformers very close in energy,
the symmetric trans structure and the a-symmetric gauche structure
(see fig.~\ref{transgauche}). The main distinction is the orientation
of the OH bond with respect to the CCO plane. A microwave
study \cite{Kakar} has shown that the trans form is slightly
(0.12~kcal/mol) more stable than the gauche form.  Table~\ref{TABLEethanol}
lists the most important geometric data of the trans and gauche
conformers, and compares the CPMD results with ADF calculations, B3LYP
calculations \cite{Borowski}, and experimental data \cite{Sasada}.  CPMD
and ADF bond lengths differ at most 0.01 \AA~ and angles are within 0.5$^o$.
Comparing CPMD with B3LYP and experimental values yields differences
upto 0.03~\AA~ and 0.5$^o$, and 0.04~\AA~ and 1$^o$ respectively.
The calculated energy difference between the two conformers is listed in
fig.~\ref{figEn}. CPMD, ADF, B3LYP of Ref.~\cite{Coussan}, and
MP4 \cite{Senent} predict the trans conformer to be stable by
0.07-0.10~kcal/mol, in good agreement with the experimental value of
$0.12$~kcal/mol. In contrast, the B3LYP calculation of \cite{Borowski}
yields the opposite, with the gauche conformer stable by $0.23$~kcal/mol. 
Vibrational frequencies obtained by a Fourier transform of the
velocity auto correlation function (VACF) of a single ethanol at 300~K
yields an OH stretch frequency of 3200 cm$^{-1}$ and a CH stretch
frequencies in the range of 2600-2800 cm$^{-1}$. This should be
compared to the experimental values of 3653 cm$^{-1}$ \cite{Coussan},
and 2800-3000 cm$^{-1}$ \cite{Coussan,Nishi}, respectively.  The
tendency of BLYP to underestimate frequencies by $\approx$~10\% is a
known feature, and also observed in BLYP calculations of water
\cite{Sprik} and methanol \cite{ErpMe01_erratum}.

For the ethanol-water complex we distinguished four complexes, with
both the ethanol trans and gauche conformer acting as proton donor or
acceptor in the hydrogen bond with water.  Table~\ref{TABLEetha-wat}
lists the most important geometric data obtained with CPMD and ADF.
Fig.~\ref{figEn} shows the energy differences for all complexes.  As
for the ethanol monomer there is excellent agreement between CPMD and
ADF results, with deviations within 0.02 \AA~ and 2$^o$ for bond
lengths and angles, and the binding energies within 0.2~kcal/mol. This
indicates a state-of-the art accuracy of the electronic-structure
method employed in CPMD.  The CPMD-BLYP calculation predicts the
complex with the water molecule donating a proton to the ethanol
gauche conformer to be the most stable, with a binding energy relative
to the isolated water and trans ethanol conformer of 4.75~kcal/mol.
The complex with the ethanol trans conformer donating a proton is less
stable by 0.82~kcal/mol.  Switching to the other ethanol conformer
within either complex destabilizes the complex by 0.07 and
0.4~kcal/mol, respectively.  The relative stability of the ethanol
donor and acceptor complexes is similar to that for the methanol-water
complex where CPMD-BLYP \cite{ErpMe01_erratum} and the
complete-basis-set MP2 estimate \cite{Kirschner} yields the methanol
acceptor complex stable by 0.74 and 0.35~kcal/mol, respectively.
There are no experimental data for the structure and energetics of the
ethanol-water dimer.  The MP2 result of Ref.~\cite{Masella} provides
the only high-level quantum mechanical study reported in literature.
This study does not distinguish between the trans and gauche
conformers. Therefore, a comparison with our results is somewhat
limited.  For the ethanol acceptor configuration the MP2 and CPMD-BLYP result
for the complexation energy are similar. However, for the ethanol
donor configuration the MP2 complexation energy is more than
1.3~kcal/mol larger than the CPMD-BLYP results.  Consequently, the MP2
calculations yield an opposite relative stability of the two
water-ethanol configurations, with an energy difference of
0.92~kcal/mol. It should be noted that in the same study the
methanol-water complex with the methanol as hydrogen bond donor is
found to be the most stable, in contradiction with the MP2 basis-set
limit result of Ref.~\cite{Kirschner}. This suggests that in the
Ref.~\cite{Masella} a limited basis set or other factors could have
lead to spurious reversal of the relative stability of the two
water-ethanol configurations.
The CPMD-BLYP hydrogen bond interaction energy for the ethanol-water
dimer is of the same order of magnitude as the CPMD-BLYP result for
the water-water \cite{Sprik}, water-methanol \cite{ErpMe01_erratum},
and methanol-methanol \cite{Handgraaf} dimer. Comparison of these
three dimers against high-level quantum chemical calculations and
experimental values indicates that BLYP underestimates the binding
energy by approximately 1~kcal/mol. For the ethanol-water dimer we
could expect a similar difference. Here we should add that going from
water, via methanol, to ethanol the dispersion forces become
increasingly important. These are not accounted for in
gradient-corrected functionals such as BLYP. Correlated methods such
MP2 incorporate, to a good approximation, dispersion forces. In
Ref.~\cite{Handgraaf} we estimated, in a comparison of BLYP and
high-level MP2 calculations \cite{Mooij} for the methanol-dimer in
non-hydrogen bonded configurations, that the absence of dispersion
interaction in BLYP amounted to an underestimate of the binding energy
of $\approx$1~kcal/mol. For the ethanol-water dimer this number could
serve as an underestimate. Yet, although by far no insignificant, the
magnitude of the deviation is much smaller than the hydrogen-bond
interaction. Therefore, it can be argued that for a study of aqueous
ethanol neglecting the dispersion interaction is acceptable.

Next we discuss the ethylene-water complexes. We will consider
complexes with a single water and a water dimer.  The dominant
interaction of water with ethylene is a $\pi$-H bond, where a proton
of the water molecule binds to $\pi$ electrons of the double C=C bond.
There exist two stable complexes between ethylene and a single water
molecule, indicated as the EW1a and EW1b geometry (see fig.
\ref{EW1aEW1b}). Both geometries have Cs symmetry with the plane of
the water molecule orthogonal to that of ethylene molecule. In the
EW1a structure the water-plane is parallel to the C=C bond, whereas in
EW1b it is orthogonal.  In table \ref{TABLEethewat} we compare the CPMD-BLYP
geometries with ADF-BLYP and MP2\cite{Tarakeshwar,Dupre} and
CCSD \cite{Tarakeshwar3} results.
Fig.~\ref{figEn} shows the energy differences, and compares these to
values reported in literature. Again the CPMD and ADF results are in
excellent agreement, with energies smaller than 0.02~kcal/mol, and
geometries within 0.01~\AA~ and 0.3$^\circ$, except for the non bonded
OH$_{nb}$ distance that is related to water orientation, a coordinate
along with the energy surface is relatively flat. The BLYP result
shows a significant $\pi$-H binding energy of ~$\approx$~1.4~kcal/mol.
However, in MP2 \cite{Tarakeshwar,Tarakeshwar3,Dupre} and 
CCSD \cite{Tarakeshwar3}
calculations, binding energies are $\approx$~1-2~kcal/mol stronger,
accompanied by a shorter $\pi$-H bond with differences of 0.12~\AA~
and 0.14~\AA~ for EW1a and EW1b, respectively.  The comparison with
the experimental value gives a similar picture with the BLYP
underestimating the binding energy $\approx$~2~kcal/mol.

Matrix isolation studies\cite{Engdahl,Engdahl86} have revealed that
the complex of ethylene with two water molecules consists of a water
dimer that has one water molecule $\pi$-H bonded to the C=C double
bond. Recent high-level MP2 calculations have indicated that the
presence of the second water molecule enhances the strength of the
$\pi$-H bond\cite{Tarakeshwar2,Dupre}. From fig.~\ref{figEn} we see
that BLYP result are qualitatively in line with this observation, with
a binding energy of the (water-dimer)-ethylene complex of
2.25~kcal/mol, up from 1.41~kcal/mol for the single-water binding.
However, quantitatively MP2 and BLYP compare less well, with BLYP
underestimating both the total binding energy as well as the increase
from the single-water binding.  Again, the experimental
value\cite{Engdahl,Engdahl86}, that is closer to the MP2 result, shows
a significantly stronger binding than BLYP.  Both for the
water-ethylene and the (water-dimer)-ethylene complex the differences
of BLYP with MP2 and experiment are significant relative to the
total binding energy. This absence of dispersion interaction in BLYP
will be an important factor contribution to this discrepancy.

Overall, we conclude that the reference calculations provide
confidence that BLYP is capable of a quantitative study of a the aqueous solvation 
of a single ethanol molecule where interactions are dominated by 
relatively strong hydrogen bonds.  BLYP qualitatively accounts for the
weaker $\pi$-H binding in water-ethylene systems.  However, comparison
with MP2 and experimental data suggest that its strength is
significantly underestimated. Yet, for a single solvated ethylene the
hydrogen bonds among the solvating water molecules will be the
dominant interaction. Hence, we believe that BLYP will be able to
quantitative described dilute aqueous ethylene solution.  Both for
ethanol and ethylene the absence of dispersion attraction in BLYP will
have some impact, in particular for the coordination around the CH$_2$
and CH$_3$ groups, where BLYP only accounts for the steric repulsion.

\section{\bf Solvation Structure}
\label{secstruc}
\subsection{Ethanol Solvation}

Fig. \ref{figrad} shows ethanol-water radial distribution functions
(RDFs) of the small and large ethanol solutions. The pronounced
structure in the hydrogen bonding RDFs (HH, OH, HO, OO) are a clear
indication of the presence of hydrogen bonds. All RDFs show that the
small system gives a good description of the first solvation shell,
while the large system also includes the second solvation shell.
The peak positions of the force-field simulations of Fidler and Rodger
\cite{Fidler}, indicated by crosses, are close to our results.
Note that the good agreement for the CO RDFs, that are potentially
sensitive to a proper description of the dispersion attraction,
suggests that the absence of the dispersion interaction in our
CPMD-BLYP simulation is of limited importance for a proper description
of the aqueous solvation of the hydrophobic group of ethanol.  The
position of the first peak of the OH- and HO-RDF indicates that the
average hydrogen-bond length is 1.7~\AA~ for both the ethanol-donor and
-acceptor bond.  Integration of the OO-RDF upto the first minimum
$r=3.3$ \AA~ yields on average 3 water molecules in the first
solvation shell of the hydroxyl group, in good agreement with the
neutron diffraction value of $\approx$~3 \cite{Sidhu} .  Integrating
the methyl-oxygen (C$_2$O) RDF up to $r=5.7$ \AA~ indicates that the
first solvation shell of the methyl group consists of approximately 21
water molecules. These coordination numbers are of the same order as
the experimental estimation of Petrillo et al.  \cite{Petrillo}, who
found that there are $18 \pm 2$ water molecules within 4 \AA~
from the center-of-mass of an ethanol in aqueous solution.

\subsection{Ethylene solvation}
In fig.~\ref{figradethe} we show the ethylene-water RDFs of the small
and large ethylene solution.  Both the carbon-oxygen and the
intermolecular carbon-hydrogen RDFs are shown.  Experimental or
simulation data for ethylene in aqueous solution at room temperature
were not available.  The comparison between the small and large system
shows a larger deviation in the first solvation shell than was found
for the ethanol solutions. Apparently, the small ethylene system is
not able to accommodate properly the first solvation shell.
Integration the CO-RDF of the large system up to the first minimum at
5.7 \AA~ yields a hydration shell of 23 water molecules.  More spatial
information can be extracted from fig. \ref{figHdist} where, for the
large system, the distribution of water H-atoms around the ethylene
C=C axis is shown.  $X$ and $Y$ are the components of the vector
joining a water hydrogen and the midpoint of the C=C bond. Here,
$X$ is the distance orthogonal to the C=C axis and $Y$ the parallel
distance.  Note that we used the four-fold symmetry to improve the
statistical accuracy yielding four identical quadrants.  The ethylene
hydrogens are shown to visualize the size of the ethylene molecule,
but do not indicate any angle dependence around the C=C axis.  The
figure shows a well defined elliptic region with no hydrogens present
except for two weak, but clearly visible, peaks in the mid plane $Y=0$
at $X=\pm 2.5$.  These should be attributed to the presence of $\pi-H$ 
bonded configurations.

\subsection{Water Structure}

In fig.~\ref{figradwater} the calculated water oxygen-oxygen RDFs of
both the ethanol and the ethylene solutions are compared to those
calculated for pure water.  Note that the first solvation shell
contains a large fraction of the total number of water molecules,
especially for the small system.  Structural changes due to the solute
molecule should therefore be detectable by comparing these systems
with the pure water system.  For the ethanol solution we see a small
drop of the first peak accompanied by a slight increase at the first
minimum.  This indicates some decrease in the hydrogen-bond structure,
when compared to the pure water.
As the small ethylene-water solution is too small to accommodate a
fully relaxed water solvation shell fig.~\ref{figradwater} only shows
the RDFs of the large ethylene system.  The RDFs show a similar
behavior as for the ethanol solvation, with a slight decrease of the
first peak and a small increase of the RDF in the region around the
first minimum. 
The small changes in the structure of the solvating water shell around
ethanol and ethylene indicate that the hydrogen-bonded water network
is very flexible and can easily open up to accommodate small apolar
solute groups without significantly changing its structure.  Our
findings are consistent with the neutron diffraction experiments of
Turner and Soper \cite{Turner} and force-field molecular dynamics
simulation of Fidler and Rodger \cite{Fidler} who also did not find
any evidence of structural enhancement in the hydration shell of
ethanol.  Turner and Soper found that only for larger hydrophobic
solutes this effect was experimentally detectable, but even then very
small.  The same trend was found for in a CPMD-BLYP molecular dynamics
study of the solvation of methanol \cite{ErpMe01_erratum}.

\subsection{Hydrogen Bonds}
To examine the hydrogen bond statistics in the ethanol solution we
adopted the definition of Ref. \cite{FeHa90}: two molecules are
hydrogen bonded if simultaneously the inter-oxygen distance is less
than 3.4 \AA~ and the OHO angle is smaller than 30$^\circ$. With this
definition we found that, for the large system, ethanol oxygen donates
on average 0.9 hydrogen bonds and accepts 1.7. For the small system we
found 1.0 and 1.6, respectively.  This in consistent with the fact
that approximately three water molecules occupy the first solvation
shell of the hydroxyl group.  For comparison, in a CPMD-BLYP
simulation we found that methanol in a dilute aqueous solution donates
on average 0.9 hydrogen bonds and accepts 1.5 \cite{ErpMe01_erratum}.
>From the pure water simulation these numbers were measured to be 1.7.

For the ethylene molecule a well defined definition for the $\pi$-H hydrogen
bond does not exist. 
To investigate the influence of the $\pi$-H hydrogen bond on the solvation 
structure we looked at the water
hydrogen positions relative to the C=C axis (see fig \ref{figHdist}).
Inside the
elliptic region around the C=C bond that is depleted of hydrogens we clearly 
detect near the center-of-mass of the ethylene molecule at a distance
of $\approx$~2.5 \AA~ a  region with an increased amount of water hydrogens.
Integration over this region with $0< X < 2.6$ and $0<Y<1$ yields a
value of 0.42. This implies that approximately 40\% of the time a
water molecule is oriented towards the double bond, forming a $\pi$-H
bond.

\section{\bf Solute Dynamics}
\label{secdyn}

The time scale of the present simulations (5-10~ps) allows for an
analysis of the short-time dynamics.  As we mentioned above BLYP
underestimates most vibrational frequencies by more than 10 \%.
However, the frequency shifts upon solvation still provide valuable
information and allow for a direct comparison to experiments.
Fig.~\ref{vibCHstretch} shows the spectrum of the velocity auto
correlation functions (VACFs) of the ethanol hydroxyl hydrogen for the
large ethanol solution and for the single 
isolated ethanol.  The spectrum is of limited accuracy due to the
relative short trajectory.  However, the calculations show that, upon
solvation, there is a clear red shift of about 200~cm$^{-1}$ of the OH
stretch frequency.  Experimental data for the OH frequency shift of
ethanol in dilute aqueous solution is not available. However, the OH red
shift is a typical characteristic of a hydrogen bonded liquid.  A
comparison with measured shifts in liquid ethanol, from
$3676~cm^{-1}$\cite{Perchard} to 3330~cm$^{-1}$\cite{Lock}, shows a
similar trend.  In a CPMD-BLYP simulation of a dilute aqueous
solution of methanol we found a similar red shift for the methanol OH
frequency of about 200~cm$^{-1}$

The vibrational spectrum of the VACF of the ethylene molecule is not
shown.  As for the ethanol system there was a limited accuracy due to
the relatively short calculated trajectory.  We observed that upon
solvation the CH and CC peaks do broaden. However an estimate for peak
shift falls outside the accuracy of the calculated spectra.
The matrix isolation studies of Engdahl and Nelander \cite{Engdahl,Engdahl86}
show minor changes in frequencies when isolated ethylene is compared
with the ethylene-water complex, with the largest shift being a blue shift
of 12 cm$^{-1}$ of the out-of-plane bending mode ($\nu_7 \approx 947$
cm$^{-1}$).

\section{\bf Polarization}
\label{secEle}
As the electronic structure is an intrinsic part of a CPMD simulation,
detailed information about the electronic charge distribution can be
obtained. To quantify the charge distribution we used the method of
maximally localized Wannier functions that transforms the
Kohn-Sham orbitals into Wannier functions, whose centers (WFC) can be
assigned a chemical meaning such as being associated with an electron
bonding- or lone-pair \cite{Marzari,Silvestrelli4}.  It also provides
a unambiguous route to determine the dipole moment of individual
molecules in a condensed phase by assuming the electronic charge to be
distributed as point charges located on the WFCs.

Table \ref{TABLEdip} list the calculated dipole moments for the
gas-phase systems and the average dipole moment for the solvated
systems. The latter were obtained from 28 independent configurations
of the small solvated systems. The distributions of these dipole moments
are shown in fig. \ref{histdip}.
For ethanol we observe an significant increase in going from the
gas-phase molecule via the water-ethanol cluster to the fully solvated
system. This trend is very similar to what is found is found for
water \cite{Silvestrelli2,Silvestrelli3} and methanol \cite{Handgraaf}
and may be considered typical feature of a strongly hydrogen bonding
molecule. The calculated distribution of dipole moments of solvated
ethanol shows that thermally driven fluctuations give rise to a
significant variation ranging from 2.0~D to 4.0~D.

Also for ethylene we observed a significant change of the dipole
moment upon solvation. Being apolar in the gas phase, the average
dipole moment increases up to 0.5~D when complexed with the water
dimer or when in solution. In solution it exhibits rare fluctuations
where the dipole moment reaches values of up to 1.0~D.
Fig.~\ref{figsnap} shows a typical snapshot where such an extreme high
value of the dipole moment is reached.
The two WFCs of the ethylene double bond are located near the middle
of the C=C bond, just below and above the plane of the molecule. In
the figure we see clearly that the upper $\pi$-WFC is acting as proton
acceptor.  The $\pi$-H bond shifts this WFC further out of plane
inducing an ethylene dipole moment orthogonal to the plane of the
molecule.  The snapshot suggests an enhancement of the ethylene
polarization by the fact that the $\pi$-H bonded water molecule has in
turn a rare four-fold 
hydrogen-bonded water coordination, with three of the waters
donating a proton.  This should be seen as a manifestation of the
strengthening of the $\pi$-H bond upon complexing water molecules with
the water molecule that is $\pi$-H bonded to the ethylene, a
feature found in gas-phase MP2 calculations \cite{Tarakeshwar2,Dupre} and also in the present BLYP calculations, and discussed above in section~\ref{secgas}.

\section{\bf Conclusions}
\label{secCon}
We have studied the solvation of ethanol and ethylene in
water by DFT-based (Car-Parrinello) Molecular Dynamics. Validation of
the computational approach by comparing structure and energetics of
relevant gas-phase complexes against experimental results and state of
the art quantum chemical calculations showed that CPMD employing the DFT-BLYP
is capable of qualitatively describing the
aqueous solvation of a single ethanol of ethylene molecule.

The structural properties of the ethanol solvation were in good
agreement with both neutron diffraction data \cite{Turner} and force-field
MD simulations  \cite{Fidler}.  We found that in
aqueous solution ethanol accepts on average 1.7 hydrogen bonds and donates on
average 0.9 hydrogen bonds.  For ethylene we found it has
approximately 0.4 $\pi$-H bonds with a water molecule.  Both for ethanol
and ethylene the simulations provided no structural evidence for
hydrophobic hydration: the structure of the hydrating water shell was
not enhanced compared to that of pure water.  The calculation even
indicated a slight decrease in the structure.  For aqueous ethanol the
calculated red shift of the hydroxyl vibration upon solvation  was
consistent with experimental findings.

Analysis of the electronic charge distribution by means of Wannier
functions showed that the interaction with water can significantly
increase the dipole moment of the ethanol and the ethylene molecule.
The average dipole moment of ethanol increases from 1.8 D in the gas
phase to 3.1 D in aqueous solution.  Ethylene, that is apolar in the
gas phase, attains an average dipole moment of 0.5~D in solution. We
identified configurations, with a $\pi$-H bonded water molecule that
has a rare four-fold hydrogen-bonded water coordination, where the
instantaneous dipole moment of ethylene takes values of up to 1~D.
Such configurations with large solute dipole moments may also play an
important role in activating chemical reactions involving the solutes
as we have seen in a CPMD-BLYP study of the acid-catalyzed hydration
of ethene \cite{ownfut}. The strong polarization effect raises
questions towards the common consideration of thinking ethylene as a
apolar and hydrophobic molecule. The electronic charge analysis also
points out the necessity of polarizable force fields for both ethanol
and ethylene when dissolved in water. The present results may be
considered valuable for the design of such force-fields.

\acknowledgments
T.S.v.E acknowledges NWO-CW (Nederlandse Organisatie voor
Wetenschappelijk Onderzoek, Chemische Wetenschappen).
E.J.M. acknowledges the Royal Netherlands Academy of Art and Sciences for
financial support.  We acknowledge support from the Stichting
Nationale Computerfacileiten (NCF) and the Nederlandse Organisatie
voor Wetenschappelijk Onderzoek (NWO) for the use of supercomputer
facilities.


\begin{thebibliography}{10}

\bibitem{FranksIves}
F. Franks and D.~J. Ives, Rev. Chem. Soc. {\bf 20},  1  (1966).

\bibitem{Eisenberg}
D. Eisenberg and W. Kauzmann, {\em The Structure and Properties of Water}
  (Clarendon Press, Oxford, U.K., 1969).

\bibitem{Franks}
F. Franks,  in {\em Water: A comprehensive treatised}, edited by F. Franks
  (Plenum Press, New York, 1973), Vol.~2.

\bibitem{FranksReid}
F.~F. D.~S. Reid,  in {\em Water: A comprehensive treatised}, edited by F.
  Franks (Plenum Press, New York, 1973), Vol.~2.

\bibitem{Blandamer}
M. Blandamer,  in {\em Water: A comprehensive treatised}, edited by F. Franks
  (Plenum Press, New York, 1973), Vol.~2.

\bibitem{Luck}
{\em Structure of Water and Aqueous Solutions}, edited by W.~A.~P. Luck (Verlag
  Chemie-Physik, Weinheim, Germany, 1974).

\bibitem{Franksv4}
F. Franks,  in {\em Water: A comprehensive treatised}, edited by F. Franks
  (Plenum Press, New York, 1975), Vol.~4.

\bibitem{Alfsen}
{\em L'eau et les syst\`emes biologiques; (Water and biological systems)},
  edited by A. Alfsen and A.~J. Berteaud (Editions due CNRS, Paris, 1976).

\bibitem{Franks2}
F. Franks and S. Mathias, {\em Biophysics of Water} (Wiley, New York, 1982).

\bibitem{FranksDes}
F. Franks and J.~E. Desnoyers,  in {\em Water science reviews}, edited by F.
  Franks (Cambridge Univ. Press, Cambridge, 1985), Vol.~2.

\bibitem{Mohr}
M. Mohr, D. Marx, M. Parrinello, and H. Zipse, Chem. Eur. J. {\bf 6},  4009
  (2000).

\bibitem{ownfut}
T.~S. van Erp and E.~J. Meijer,   (to be published).

\bibitem{Coccia}
A. Coccia, P.~L. Indovina, F. Pado, and V. Viti, Chem. Phys. {\bf 7},  30
  (1975).

\bibitem{Harris}
K.~R. Harris, P.~J. Newitt, and Z.~J. Derlacki, J. Chem. Soc. Faraday Trans.
  {\bf 94},  1963  (1998).

\bibitem{Mizuno}
K. Mizuno, Y. Miyashita, Y. Shindo, and H. Ogawa, J. Phys. Chem. {\bf 99},
  3225  (1995).

\bibitem{Ludwig}
R. Ludwig, Chem. Phys. {\bf 195},  329  (1995).

\bibitem{Lamanna}
R. Lamanna and S. Cannistraro, Chem. Phys. {\bf 213},  95  (1996).

\bibitem{Brai}
M. Brai and U. Kaatze, J. Phys. Chem. {\bf 96},  8946  (1992).

\bibitem{Nishi}
N. Nishi, S. Takahashi, M. Matsumoto, A. Tanaka, K. Muraya, T. Takamuku, and T.
  Yamaguchi, J. Phys. Chem. {\bf 99},  462  (1995).

\bibitem{Onori}
G. Onori and A. Santucci, J. Mol. Liq. {\bf 69},  1661  (1996).

\bibitem{Nishi0}
N. Nishi, K. Koga, C. Oshima, K. Yamamoto, U. Nagashima, and K. Nagami, J. Am.
  Chem. Soc. {\bf 110},  5246  (1988).

\bibitem{Petrillo}
C. Petrillo, G. Onori, and F. Sacchetti, Mol. Phys. {\bf 67},  697  (1989).

\bibitem{Turner}
J. Turner and A.~K. Soper, J. Chem. Phys. {\bf 101},  6116  (1994).

\bibitem{Sidhu}
K.~S. Sidhu, J.~M. Goodfellow, and J.~Z. Turner, J. Chem. Phys. {\bf 110},
  7943  (1999).

\bibitem{Mashimo}
S. Mashimo, T. Umehara, and H. Redlin, J. Chem. Phys. {\bf 95},  6257  (1991).

\bibitem{Bao}
J.~Z. Bao, M.~L. Swicord, and C.~C. Davis, J. Chem. Phys. {\bf 104},  4441
  (1996).

\bibitem{Sato}
T. Sato, A. Chiba, and R. Nozaki, J. Chem. Phys. {\bf 110},  2508  (1999).

\bibitem{Petong}
P. Petong, R. Pottel, and U. Kaatze, J. Phys. Chem. A {\bf 104},  7420  (2000).

\bibitem{Mueller-Plathe}
F. M\"{u}eller-Plathe, Mol. Sim. {\bf 18},  133  (1996).

\bibitem{Levchuk}
V.~N. Levchuk, I.~I. Sheykhet, and B.~Y. Simkin, Chem. Phys. Lett. {\bf 185},
  339  (1991).

\bibitem{Fidler}
J. Fidler and P.~M. Rodger, J. Phys. Chem. B {\bf 103},  7695  (1999).

\bibitem{Tarek}
M. Tarek, D.~J. Tobias, and M.~L. Klein, . Chem. Soc., Faraday Trans. {\bf 92},
   559  (1996).

\bibitem{Tarek2}
M. Tarek, D.~J. Tobias, and M.~L. Klein, Physica A {\bf 231},  117  (1996).

\bibitem{Frank}
H.~S. Frank and M.~W. Evans, J. Chem. Phys. {\bf 13},  507  (1945).

\bibitem{Ben-Naim}
A. Ben-Naim, {\em Hydrophobic Interactions} (Plenum Press, New York, 1980).

\bibitem{Sugahara}
T. Sugahara, K. Morita, and K. Ohgaki, Chem. Eng. Science {\bf 55},  6015
  (2000).

\bibitem{Tanaka}
H. Tanaka, Fluid Phase Equilibria {\bf 144},  361  (1998).

\bibitem{Kvamme}
B. Kvamme and H. Tanaka, J. Phys. Chem. {\bf 99},  7114  (1995).

\bibitem{DelBene}
J.~E. {Del~Bene}, Chem. Phys. Lett. {\bf 24},  203  (1974).

\bibitem{Engdahl}
A. Engdahl and B. Nelander, Chem. Phys. Lett. {\bf 113},  49  (1985).

\bibitem{Engdahl86}
A. Engdahl and B. Nelander, J. Phys. Chem. {\bf 90},  4982  (1986).

\bibitem{Andrews}
A.~M. Andrews and R.~L. Kuczkowski, J. Chem. Phys. {\bf 98},  791  (1993).

\bibitem{Peterson}
K.~I. Peterson and W. Klemperer, J. Chem. Phys. {\bf 85},  725  (1986).

\bibitem{Tarakeshwar}
P. Tarakeshwar, H.~S. Choi, S.~J. Lee, J.~Y. Lee, K.~S. Kim, T.-K. Ha, J.~H.
  Jang, J.~G. Lee, and H. Lee, J. Chem. Phys. {\bf 111},  5838  (1999).

\bibitem{Tarakeshwar2}
P. Tarakeshwar, K.~S. Kim, and B. Brutschy, J.~Chem.~Phys. {\bf 112},  1769
  (2000).

\bibitem{Tarakeshwar3}
P. Tarakeshwar, H.~S. Choi, and K.~S. Kim, J.~Am.~Chem.~Soc. {\bf 123},  3323
  (2001).

\bibitem{Dupre}
D.~B. DuPr\'{e} and M.~C. Yappert, J. Phys. Chem. A {\bf 106},  567  (2002).

\bibitem{CaPa85}
R. Car and M. Parrinello, Phys. Rev. Lett. {\bf 55},  2471  (1985).

\bibitem{Laasonen}
K. Laasonen, M. Sprik, M. Parrinello, and R. Car, J. Chem. Phys. {\bf 99},
  9080  (1993).

\bibitem{Sprik}
M. Sprik, J. Hutter, and M. Parrinello, J. Chem. Phys. {\bf 105},  1142
  (1996).

\bibitem{SiBe97}
P.~L. Silvestrelli, M. Bernasconi, and M. Parrinello, Chem. Phys. Lett. {\bf
  277},  478  (1997).

\bibitem{Marx}
D. Marx, M. Sprik, and M. Parrinello, Chem. Phys. Lett. {\bf 273},  360
  (1997).

\bibitem{ErpMe01_erratum}
T.~S. van Erp and E.~J. Meijer, Chem. Phys. Lett. {\bf 333},  290  (2001), (by
  mistake the labels (a) and (b) in figure 1 are reversed).

\bibitem{Raugei}
S. Raugei and M.~L. Klein, J. Chem. Phys. {\bf 116},  196  (2002).

\bibitem{KoSh65}
W. Kohn and L.~J. Sham, Phys. Rev. {\bf 140},  1133  (1965).

\bibitem{HoKo64}
P. Hohenberg and W. Kohn, Phys. Rev. B {\bf 136},  864  (1964).

\bibitem{Beck88_2}
A.~D. Becke, Phys. Rev. A {\bf 38},  3098  (1988).

\bibitem{LeYa88}
C. Lee, W. Yang, and R.~G. Parr, Phys. Rev. B {\bf 37},  785  (1988).

\bibitem{CPMD30f}
CPMD, version 3.0f, developed by J. Hutter, P. Ballone, M. Bernasconi, P.
  Focher, E. Fois, S. Goedecker, M. Parrinello, and M. Tuckermann, at MPI f\"ur
  Festk\"orperforschung and IBM Zurich Research Laboratory (1990-1997).

\bibitem{TrMa91}
N. Troullier and J.~L. Martins, Phys. Rev. B {\bf 43},  1993  (1991).

\bibitem{ADF2002}
ADF~2002.01, SCM, Theoretical Chemistry, Vrije Universiteit, Amsterdam, The
  Netherlands, http://www.scm.com.

\bibitem{Velde}
G. te~Velde, F. Bickelhaupt, E. Baerends, G.~F. Guerra, S. van Gisbergen, J.
  Snijders, and T. Ziegler, J. Comput. Chem. {\bf 22},  931  (2001).

\bibitem{Fonseca}
C.~F. Guerra, J.~G. Snijders, G. te~Velde, and E.~J. Baerends, Theor. Chem.
  Acc. {\bf 99},  391  (1998).

\bibitem{ADF_basisVII}
Kohn-Sham orbitals are expanded in an even-tempered, all-electron Slater type
  basis set augmented with 2p and 3d polarization functions for H and 3d and 4f
  polarization functions for C and O.

\bibitem{BaLa93}
R.~N. Barnett and U. Landman, Phys. Rev. B {\bf 48},  2081  (1993).

\bibitem{Nose84_1}
S. Nos\'e, J. Chem. Phys. {\bf 81},  511  (1984).

\bibitem{Nose84_2}
S. Nos\'e, Mol. Phys. {\bf 52},  255  (1984).

\bibitem{Hoov85}
W.~G. Hoover, Phys. Rev. A {\bf 31},  1695  (1985).

\bibitem{Kakar}
R.~K. Kakar and C.~R. Quade, J. Chem. Phys. {\bf 72},  4300  (1980).

\bibitem{Borowski}
P. Borowski, Janowski, and Woli\'nski, Mol. Phys. {\bf 98},  1331  (2000).

\bibitem{Sasada}
Y. Sasada, M. Takano, and T. Satoh, Mol. Spec. {\bf 38},  33  (1971).

\bibitem{Coussan}
S. Coussan, Y. Bouteiller, J.~P. Perchard, and W.~Q. Zheng, J. Phys. Chem. A
  {\bf 102},  5789  (1998).

\bibitem{Senent}
M.~L. Senent, Y.~G. Smeyers, R. Dom\'inguez-G\'omez, and M. Villa, J. Chem.
  Phys. {\bf 112},  5809  (2000).

\bibitem{Kirschner}
K.~N. Kirschner and R.~J. Woods, J. Phys. Chem. A {\bf 105},  4150  (2001).

\bibitem{Masella}
M. Masella and J.~P. Flament, J. Chem. Phys. {\bf 108},  7141  (1998).

\bibitem{Handgraaf}
J.~W. Handgraaf, T.~S. van Erp, and E.~J. Meijer, Chem. Phys. Lett. {\bf 367},
  617  (2003).

\bibitem{Mooij}
W.~T.~M. Mooij, F.~B. van Duijneveldt, J.~G.~C.~M. van Duijneveldt-van~de
  Rijdt, and B.~P. van Eijck, J. Phys. Chem. A {\bf 103},  9872  (1999).

\bibitem{FeHa90}
M. Ferrario, M. Haughney, I.~R. McDonald, and M.~L. Klein, J. Chem. Phys. {\bf
  93},  5156  (1990).

\bibitem{Perchard}
J.~P. Perchard and M.-L. Josien, J. Chim. Phys. Phys.-Chim. Biol. {\bf 105},
  1238  (1986).

\bibitem{Lock}
S.~W. A.~J.~Lock and H.~J. Bakker, J. Phys. Chem. A. {\bf 105},  1238  (2001).

\bibitem{Marzari}
N. Marzari and D. Vanderbilt, Phys. Rev. B {\bf 56},  12847  (1997).

\bibitem{Silvestrelli4}
P.~L. Silvestrelli, N. Marzari, D. Vanderbilt, and M. Parrinello, Solid State
  Commun. {\bf 107},  7  (1998).

\bibitem{Silvestrelli2}
P.~L. Silvestrelli and M. Parrinello, Phys. Rev. Lett. {\bf 82},  3308  (1999).

\bibitem{Silvestrelli3}
P.~L. Silvestrelli and M. Parrinello, J. Chem. Phys. {\bf 111},  3572  (1999).

\end{thebibliography}

\clearpage
\begin{widetext}

{\bf \centerline{TABLES}}
\begin{table}[hb!]
\caption{Gas-phase complexes: ethanol gauche and trans monomer.
Distances in \AA, angles in degrees.
Our results: CPMD-BLYP and ADF-BLYP,
are compared with B3LYP/6-311G(2d,2p) \cite{Coussan} and experimental
microwave spectroscopy data \cite{Sasada}.}
\begin{tabular}{r|cccc}
Geometry & Gauche &&&\\ \hline
Method                    & CPMD  & ADF$$ & B3LYP  & EXP \\ \hline
$r_\textrm{OH}$           & 0.981 & 0.971 & 0.961  & 0.945 \\
$r_\textrm{CO}$           & 1.455 & 1.447 & 1.429  & 1.427 \\
$r_\textrm{CC}$           & 1.529 & 1.532 & 1.521  & 1.530  \\
$<r_\textrm{CH}>$ (in CH3)& 1.099 & 1.097 & 1.091  & 1.094 \\
$<r_\textrm{CH}>$ (in CH2)& 1.100 & 1.099 &  1.092 & 1.094 \\
$\alpha_\textrm{COH}$     & 108.0 & 108.2 & 108.7  & 108.3  \\
$\alpha_\textrm{CCO}$    & 113.0 & 113.1 & 113.0 & 112.2 \\ \hline
Geometry & Trans &&& \\ \hline
Method                    &  CPMD & ADF   & B3LYP  & EXP \\ \hline
$r_\textrm{OH}$           & 0.980 & 0.970 & 0.960  & 0.945\\
$r_\textrm{CO}$           & 1.458 & 1.449 & 1.432  & 1.425 \\
$r_\textrm{CC}$           & 1.523 & 1.524 & 1.515  & 1.530 \\
$<r_\textrm{CH}>$ (in CH3)& 1.098 & 1.096 & 1.090  & 1.094\\
$<r_\textrm{CH}>$ (in CH2)& 1.103 & 1.102 &  1.095 & 1.094 \\
$\alpha_\textrm{COH}$     & 108.4 & 108.4 & 109.0  & 108.3  \\
$\alpha_\textrm{CCO}$     & 107.6 & 108.0 & 108.0  & 107.2\\ \hline
\end{tabular}
\label{TABLEethanol}
\end{table}


\begin{table}
\caption{Gas-Phase complexes: ethanol-water dimers.
Distances in \AA, angles in degrees.
CPMD-BLYP and ADF-BLYP are compared. We differentiate four complexes with
the ethanol can be in the gauche or trans geometry and acts
as proton acceptor or proton donor in the ethanol-water hydrogen bond.
${\textrm{H}_{nb}}$ is the hydrogen of the water that is not involved in a hydrogen bond,
$\textrm{C}_1$ is the hydroxyl carbon, 
$\textrm{C}_2$ is the methyl carbon and 
$\textrm{O}_w$ is the water oxygen.}
\begin{tabular}{r|cc|cc|cc|cc|}
Geometry                       & Gauche &       &       &       & Trans  &       &       &       \\ \hline
H-bond                         & H-acc  &       & H-don &       & H-acc  &       & H-don &       \\ \hline
Method                         & CPMD   & ADF   & CPMD  & ADF   & CPMD   & ADF   & CPMD  & ADF   \\ \hline
$r_\textrm{OH}$ in H2O         & 0.994  & 0.981 & 0.981 & 0.972 & 0.993  & 0.982 & 0.981 & 0.972 \\
$r_{\textrm{OH}_{nb}}$ in H2O  & 0.981  & 0.970 & 0.981 & 0.971 & 0.978  & 0.969 & 0.982 & 0.971 \\
$r_\textrm{OH}$ in eth.        & 0.981  & 0.971 & 0.987 & 0.978 & 0.982  & 0.970 & 0.988 & 0.977 \\
$r_\textrm{OH}$ H-bond         &1.934   & 1.912 & 1.988 & 2.011 & 1.928  & 1.929 & 1.960 & 1.993 \\
$r_\textrm{CC}$                & 1.526  & 1.528 & 1.532 & 1.533 & 1.521  & 1.522 & 1.523 & 1.526 \\
$r_\textrm{CO}$                &  1.469 & 1.457 & 1.452 & 1.440 & 1.474  & 1.460 & 1.454 & 1.442 \\
$r_\textrm{OO}$                & 2.924  & 2.884 & 2.978 & 2.983 & 2.908  & 2.989 & 2.988 & 2.968 \\
$r_{\textrm{O}_w\textrm{C}_1}$ & 3.653  & 3.626 & 3.773 & 3.793 & 3.641  & 3.641 & 3.815 & 3.764 \\
$r_{\textrm{O}_w\textrm{C}_2}$ & 3.932  & 3.920 & 4.128 & 4.033 & 4.013  & 4.008 & 5.229 & 5.189 \\
$\alpha$ HOH in H2O            & 104.2  & 105.0 & 104.6 & 104.9 & 104.6  & 105.0 & 104.2 & 104.9 \\
$\alpha$ COH                   & 107.9  & 108.4 & 108.5 & 108.7 & 107.9  & 108.2 & 108.9 & 108.5 \\
$\alpha$ OH--O                 & 172.6  & 170.2 & 172.4 & 172.8 & 168.1  & 168.4 & 175.0 & 174.1 \\ \hline
\end{tabular}
\label{TABLEetha-wat}
\end{table}


\begin{table}[!hb]
\caption{Gas-Phase complexes: ethylene-water, EW1a, EW1b structures 
(see fig.~\ref{EW1aEW1b}).
Distances in \AA, angles in degrees.
CPMD-BLYP, ADF-BLYP and MP2-TZ2P++ \cite{Tarakeshwar} calculations  
are compared.
H is the water-hydrogen involved in the $\pi$-H bond; 
H$_{nb}$ is the other hydrogen of the water molecule.
CM is the midpoint of the two carbons of the ethylene molecule.}
 \begin{tabular}{r|ccc|ccc}
Geometry                  & EW1a    &          &          &EW1b    &          &       \\ \hline
Method                    &  CPMD   & ADF      & MP2      &  CPMD  & ADF      & MP2   \\ \hline
$r_\textrm{CC}$           & 1.334   & 1.336    &          &  1.335 & 1.336    &         \\ 
$r_\textrm{H-CM}$         & 2.483   & 2.485    & 2.363    & 2.570  & 2.524    &2.383    \\
$r_\textrm{O-CM}$         & 3.452   & 3.460    & 3.301    & 3.527  & 3.497    &3.337     \\
$r_\textrm{H$_{nb}$-CM}$  & 3.909   & 3.833    & 3.816    & 3.957  & 3.881    &3.793     \\
$r_\textrm{OH}$           & 0.984   &0.975     &  0.962   &0.978   & 0.974    &0.962     \\
$r_\textrm{OH$_{nb}$}$    & 0.980   & 0.970    &  0.958   &0.976   & 0.970    &0.958     \\
$\alpha_\textrm{HOH}$              & 104.2   & 104.5    &  104.7   &  104.4 &104.5     &104.7      \\
\hline
\end{tabular}
\label{TABLEethewat}
\end{table}


\begin{table}
\caption{Dipole moments obtained by Wannier-function analysis. The liquid water value
(a) was taken from Silvestrelli et al. \cite{Silvestrelli2,Silvestrelli3}.
In the last column the total dipole moments of the gas-phase complexes are given.
The first value for water
in the E2W complex is the water closest to the ethylene, 
the second value is the dipole of the other water molecule.
The solvated ethanol and ethylene dipole moments are obtained by
taking the average of 28 independent configurations from the calculated trajectories.}
 \begin{tabular}{|r|ccc|}
\hline
complex        & ethanol/ethylene &  water& total            \\ \hline
single-water   & -              & 1.82        & 1.82       \\
liquid water    & -             &3.0  (a)      &                 \\ \hline
ethanol gauche  & 1.83           & -           & 1.83       \\
ethanol trans  & 1.66           & -           &  1.66      \\
gauche P-acc   & 2.32           & 2.16        & 2.76       \\
gauche P-don   & 1.91           & 1.91        &2.62        \\
trans P-acc    & 2.13           & 2.13        & 2.47       \\
trans P-don    & 1.91           & 2.12        & 2.29       \\
solvated ethanol&  3.08 & 2.97 &                 \\ \hline
ethylene         & 0.00           & -           & 0.00       \\
EW1a           & 0.37           & 1.95        & 2.10       \\
EW1b           & 0.33           & 1.90        & 2.02       \\
E2W            & 0.51           & 2.25 , 2.17 &2.41        \\
solvated ethylene &0.51  & 3.00 &                 \\ \hline
\end{tabular}
\label{TABLEdip}
\end{table}

\clearpage
{\bf \centerline{FIGURES}}
\begin{figure}[h!]
\includegraphics[angle=0, width=8cm]{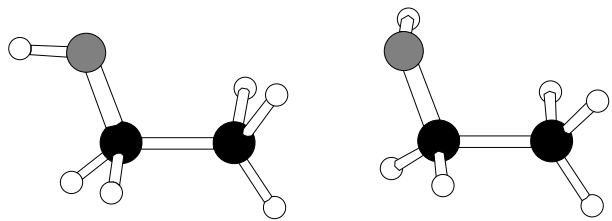}
\caption{Illustration of the trans and gauche conformers of ethanol.}
\label{transgauche}
\end{figure}

\begin{figure}[h!]
\includegraphics[angle=0, width=8cm]{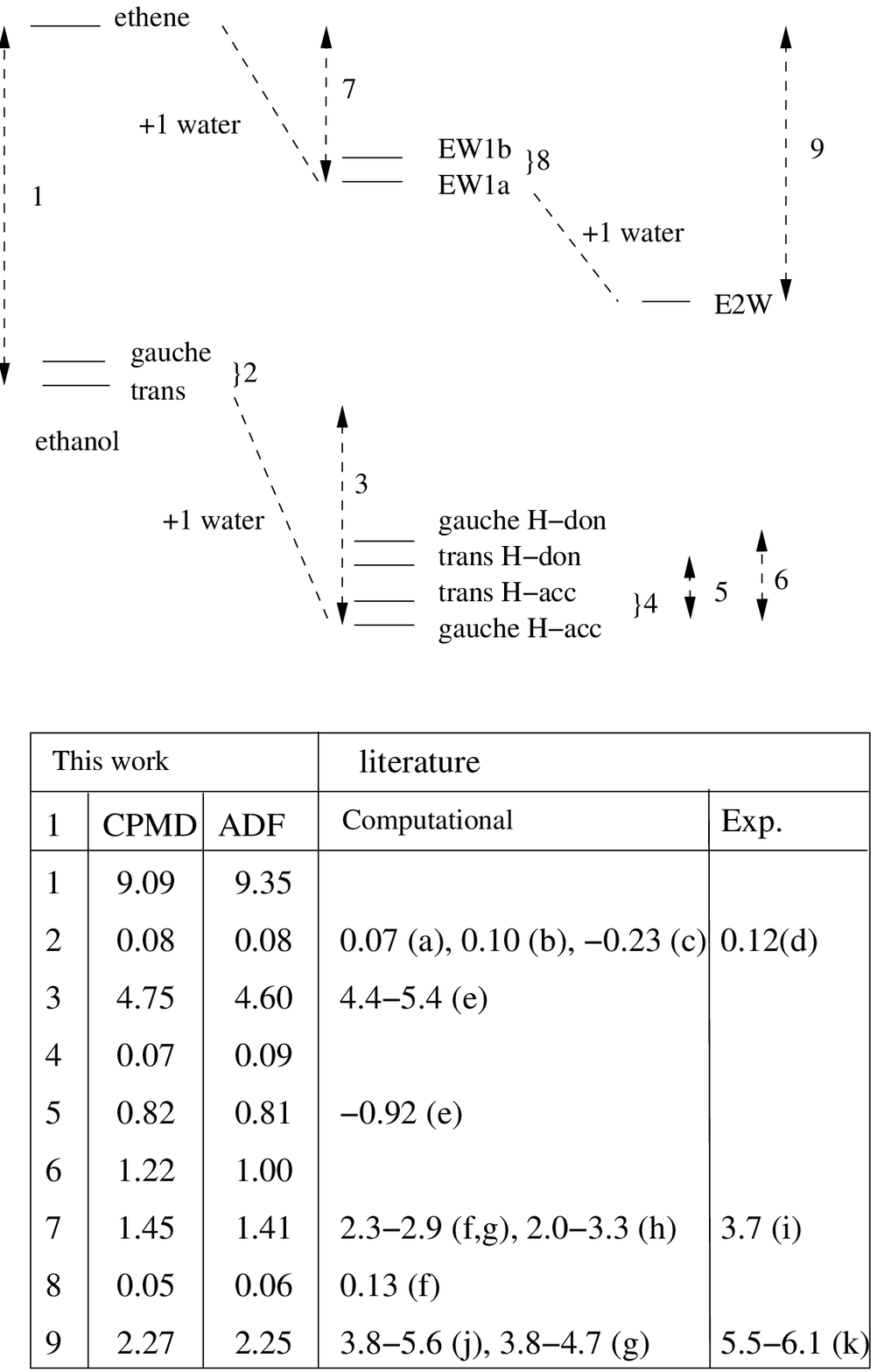}
\caption{ 
  Relative
  energies of ethylene-water and ethanol-water in kcal/mol.
  Zero-point energies (ZPE) are not included. For ethylene, the energies are
  relative to the separate molecules, whereas for ethanol the separate
  water and trans ethanol conformer is taken as the reference value.
    3 and 7 indicate the binding
  energies of ethanol and ethylene respectively.  9 is the binding
  energy between ethylene and the water-dimer. 
  The literature values are: 
  a:~MP4-(SDQT)/cc-pVTZ \cite{Senent}, 
  b:~DFT/B3LYP/6-311G(d,p) \cite{Coussan}, 
  c:~DFT/B3LYP/6-311G(d,p) \cite{Borowski},
  d: microwave experiment \cite{Kakar}
  e:~MP2/6-311+G(2df,2p)+BSSE \cite{Masella}, 
  f:~MP2/TZ2P++ \cite{Tarakeshwar}, 
  g:~MP2/6-311+G(2d,2p) \cite{Dupre},
  h:~CCSD(T)/aug-cc-pVDZ \cite{Tarakeshwar3}, 
  i:~Matrix isolation study \cite{Engdahl86}, 
  j:~MP2/aug-cc-pVDZ \cite{Tarakeshwar2}, and
  k:~Matrix isolation study \cite{Engdahl}. 
  For (e), (f), (g), and (h) we gave
the BSSE-corrected (lowest value) and the non-BSSE corrected values 
(highest value). (i) and (k) are the experimental values minus the ZPE
of \cite{Dupre}.
 }
\label{figEn}
\end{figure}

\begin{figure}[h!]
\includegraphics[angle=0, width=8cm]{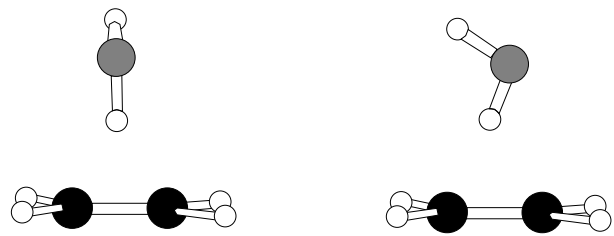}
\caption{Illustration of the EW1a and EW1b ethylene-water complexes.}
\label{EW1aEW1b}
\end{figure}

\begin{figure}[h!]
\includegraphics[width=12cm]{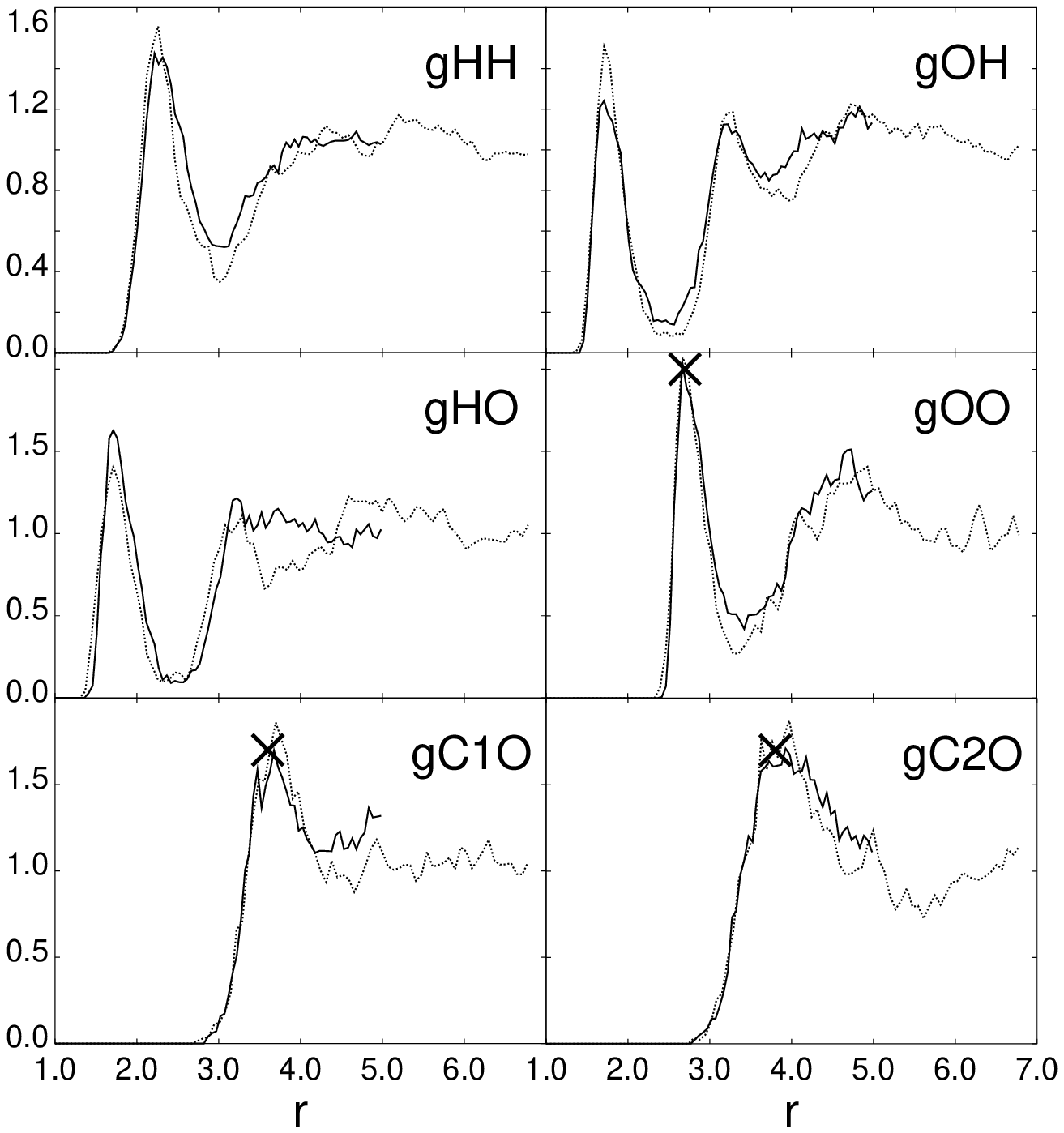}
\caption{Ethanol-water RDFs of the ethanol solutions.
  All insets show both the results of the small (solid
  line) and the large system (dashed line). For all graphs $gAB$
  indicates that the first atom $A$ belongs to ethanol and the second
  atom $B$ belongs to water.  If the first is a hydrogen, $A$=H, then
  always the hydroxyl hydrogen of the ethanol is meant. C$_1$ is the carbon bonded to
  the hydroxyl group, C$_2$ is the carbon of the methyl group.}
\label{figrad}
\end{figure}

\begin{figure}[h!]
\includegraphics[width=12cm]{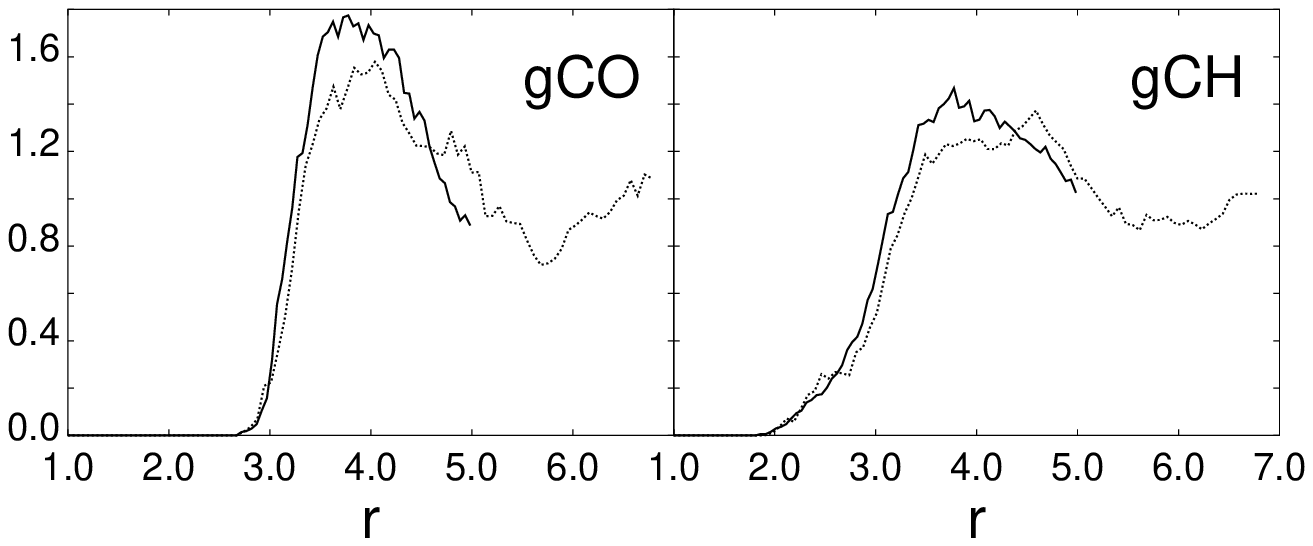}
\caption{Ethylene-water RDFs of the ethylene solutions.
The result for small system and large system are given by the solid
and dashed line, respectively. H denotes the water hydrogens only.}
\label{figradethe}
\end{figure}

\begin{figure}[h!]
\includegraphics[angle=0, width=12cm]{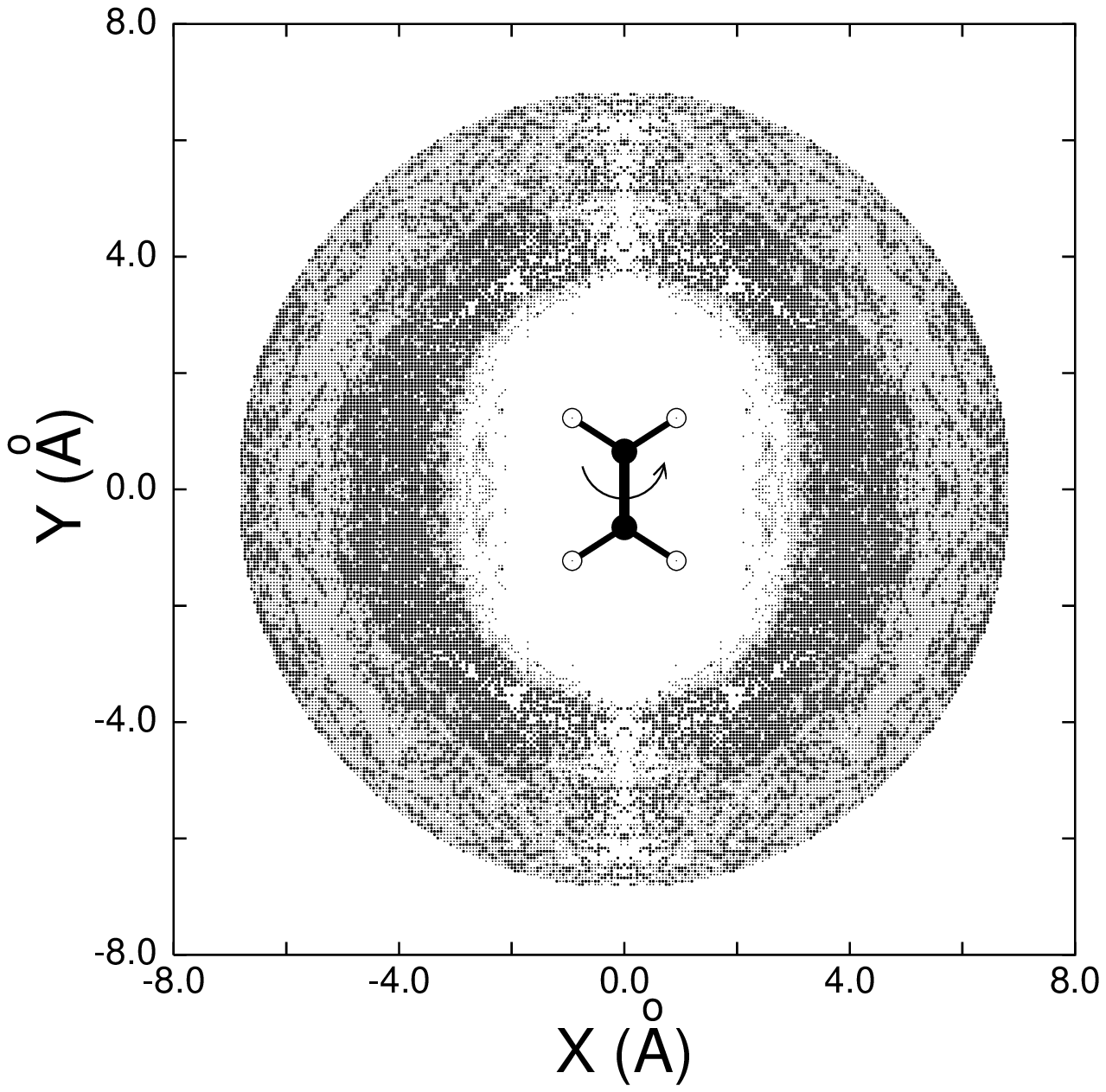}
\caption{ 
 Time and rotational averaged normalized density distribution 
  $\rho(r)/\rho_0$ (with $\rho_0$ is the average hydrogen density) of
  the water-hydrogens around the solvated ethylene molecule obtained
  from the large system simulation. The rotational average is about
  the C=C axis. The four-fold symmetry is imposed so that all
  quadrants contain the same information.
 The density regions $1.0 <
  \rho(r)/\rho_0 < 2.0$ and $ \rho(r)/\rho_0 > 2.0$ are illustrated by
  small and large pixels, respectively.}
\label{figHdist}
\end{figure}

\begin{figure}[h!]
\includegraphics[width=12cm]{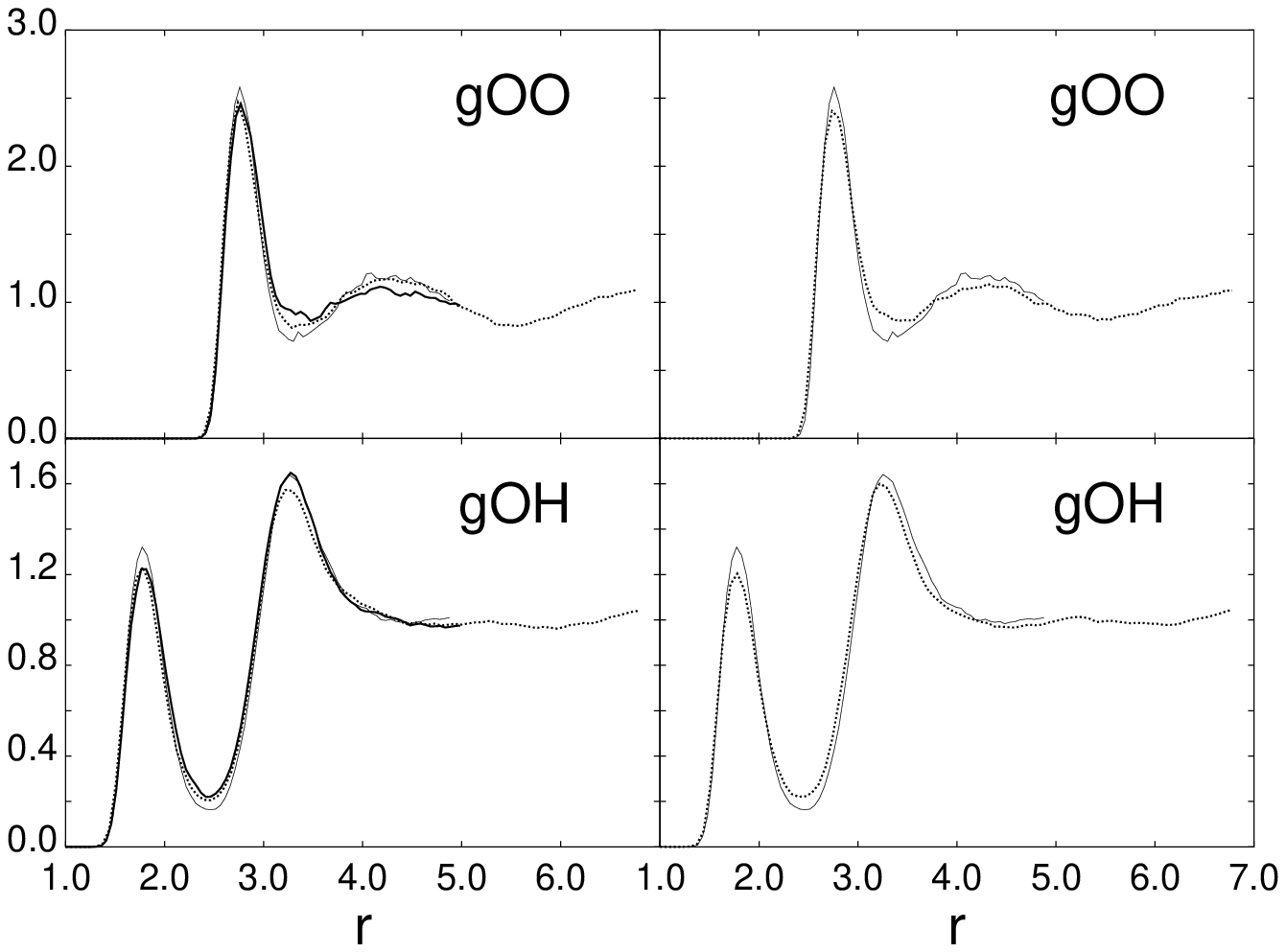}
\caption{Water-water RDFs. Left: the two ethanol solution systems are compared
  with pure water. Right: the large ethylene solution system is
  compared with pure water. Thin solid lines indicate the pure water,
  thick solid lines the small system, and thick dashed lines
  the large system.}
\label{figradwater}
\end{figure}

\begin{figure}[h!]
\includegraphics[angle=-90, width=12cm]{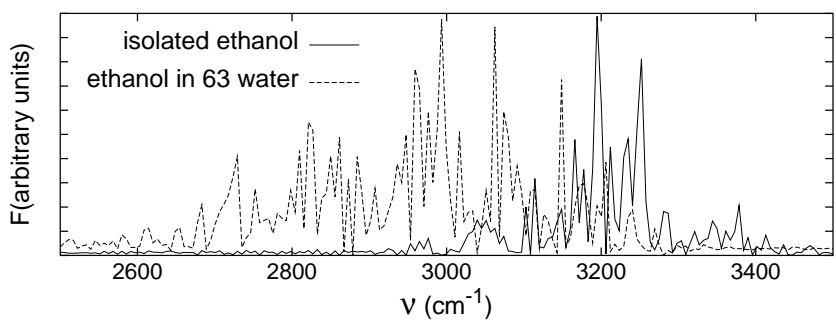}
\caption{
Spectrum of the VACF of the hydroxyl 
  ethanol hydrogens in the gas-phase
  and in the large aqueous system.}
\label{vibCHstretch}
\end{figure}

\begin{figure}[h!]
\includegraphics[angle=-90, width=10cm]{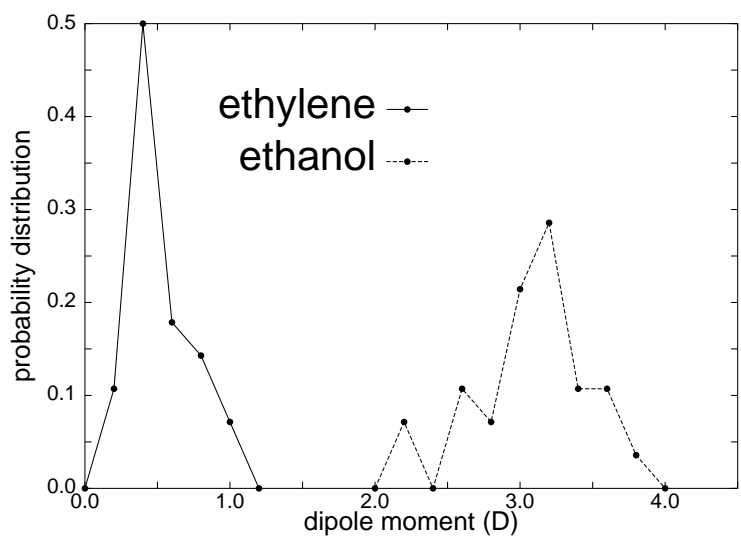}
\caption{Dipole moment distribution of the solvated ethylene and the solvated ethanol molecule.}
\label{histdip}
\end{figure}

\begin{figure}[h!]
\includegraphics[angle=0, width=9.5cm]{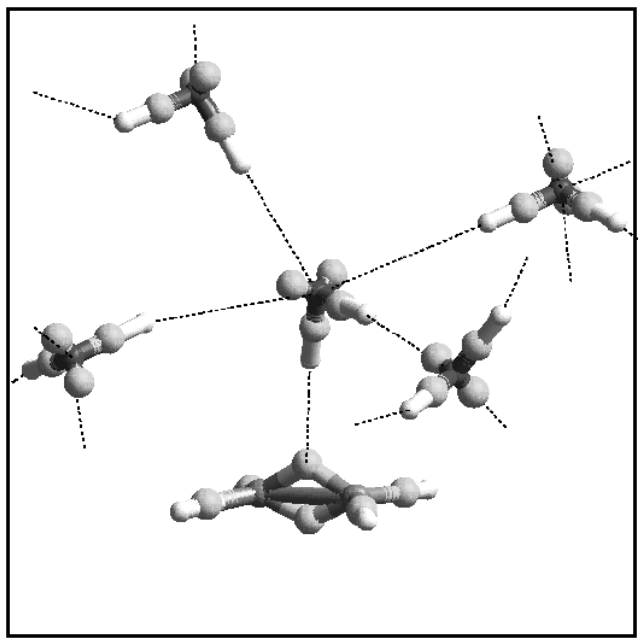}
\caption{A snapshot of  ethylene
  solvated in water. The figure shows a rare configuration where the
  ethylene molecule has a large dipole moment of $\approx$~1~D. For
  clarity, only the ethylene molecule and five water molecules are
  shown. Carbon and oxygens are dark grey.  Hydrogens are light grey.
  Besides the atoms also the WCFs are shown (middle grey).  The water
  molecules have WFCs in both OH bonds and two aside the oxygen
  corresponding to the lone pairs.  Ethylene has four WFCs along the
  CH bonds and two out-of-plane WFCs between the carbons corresponding
  to the $\pi$ bond.  The dashed lines indicate the hydrogen bonds.
  One the protons of the central water molecules points towards the
  'upper' $\pi$ bond WFC. This is the manifestation of the $\pi$-H
  bond in this picture.
}
\label{figsnap}
\end{figure}
\end{widetext}
\end{document}